\documentstyle[eqsecnum,aps,prb,twocolumn]{revtex}
\input{psfig}

\begin{document}
\draft
\twocolumn[
\hsize\textwidth\columnwidth\hsize
\csname@twocolumnfalse\endcsname

\title{Fermi surface origin of the interrelation between 
       magnetocrystalline anisotropy and compositional 
       order in transition metal alloys}

\author{S.S.A. Razee and J.B. Staunton}
\address{Department of Physics, University of Warwick,
          Coventry CV4 7AL, United Kingdom}
\author{B. Ginatempo and E. Bruno}
\address{Dipartimento di Fisica and Unit\`a INFM,
         Universit\`a di Messina, 
         Salita Sperone 31, I-98166 Messina, Italy}
\author{F.J. Pinski}
\address{Department of Physics, University of Cincinnati,
         Ohio 45221, USA}

\date{\today}

\maketitle

\begin{abstract}
Recently, we outlined a scheme to investigate the effects 
of compositional order on the magnetocrystalline anisotropy of
alloys from a first-principles electronic structure point of
view \{Phys. Rev. Lett. {\bf 83}, 5369 (1999)\} and showed that
compositional order enhances the magnitude of magnetocrystalline
anisotropy energy (MAE) of Co$_{0.5}$Pt$_{0.5}$ alloy by some
two orders of magnitude as well as affecting the equilibrium 
magnetization direction. Here we describe our scheme in
detail and present an in-depth study of the effect by
demonstrating its Fermi surface origin. In 
Co$_{0.25}$Pt$_{0.75}$ alloy we find that the perfect $ L1_2 $
structure has a very small MAE whereas imposition of directional
order enhances the MAE by two orders of magnitude.
We also present the effect
of lattice distortion (tetragonalization) on the MAE on the 
same footing and find that in the Co$_{0.5}$Pt$_{0.5}$ alloy it
accounts for only about 20\% of the observed MAE, thus 
confirming that compositional order is the major player in the 
enhancement of MAE. We also examine the directional chemical 
order that can be produced by magnetic annealing within the 
same framework. We extract a Fermi surface mechanism for the
effect in an explicit study of permalloy. Finally, we propose
that the Fermi surface plays a major role in the strong 
coupling between magnetocrystalline anisotropy and compositional
order in many magnetic alloys.
\end{abstract}

\pacs{PACS numbers: 75.30.Gw, 75.50.Cc, 75.60.Nt, 75.50.Ss}
]

\section{INTRODUCTION}
In recent years, there has been intensive theoretical as well
as experimental research on the magnetocrystalline anisotropy
of ferromagnetic materials containing transition metals, in
particular, in the form of multilayers and thin films,
because of the technological implications for high-density
magneto-optical storage media.\cite{lmf,tw,bh,mtj} Areal
densities of information storage systems are increasing
continuously and are expected to reach 40 Gbits per square inch
by the year 2004 which requires the grain size to be less than 
10 nm. For such applications the films and 
multilayers need to exhibit a very strong perpendicular 
magnetic anisotropy \cite{mhk,dnl} (PMA) to avoid 
destabilization of the magnetization of recording bits by
thermal fluctuations and demagnetizing fields.\cite{ct}
Whereas in ultrathin films and multilayers the PMA is due to 
surface \cite{xlc,had} and interface \cite{ghod,pb} effects 
respectively, in thicker films of transition metal alloys it 
is the strong intrinsic bulk magnetocrystalline anisotropy 
which leads to PMA. Thick films of transition metal
alloys are particularly interesting for magneto-optic recording 
because in addition to the required magneto-optic properties
these are chemically stable and easy to manufacture. To design 
magnetic materials for future magneto-optic recording 
applications a detailed understanding of the mechanism of 
magnetocrystalline anisotropy is needed. A significant effort 
has been directed towards an understanding of the microscopic 
mechanism of magnetocrystalline anisotropy of these systems 
from a first-principles electronic structure point of view but 
since magnetocrystalline anisotropy arises from spin-orbit 
coupling, which is essentially a relativistic 
effect \cite{hb,hjfj} this means that a fully relativistic 
electronic structure framework is desirable.

Several experimental observations indicate a correlation
between the compositional order and the magnetocrystalline
anisotropy of ferromagnetic alloys, both in the 
bulk \cite{gh,bz,ram,nm1,nm2,nm3,sc,etf} as well as in 
films.\cite{dw1,dw2,mm1,mm2,wg1,si1,si2,grh,yy,sh,tat,pwr,pk1}
Very recently,\cite{our3} we developed a `first-principles'
theory of the interrelationship between the magnetocrystalline
anisotropy and compositional order, both short- and
long-ranged. In this theory the electronic structure is treated
within the spin-polarized fully relativistic
Korringa-Kohn-Rostoker coherent-potential 
approximation \cite{he} (SPR-KKR-CPA) and the compositional 
order is modeled using the framework of static concentration
waves \cite{agk} and is an extension of our earlier
works \cite{our1,our2} on the magnetocrystalline anisotropy
of disordered alloys. Within this scheme we showed for the
first time that in fcc-Co$_{0.5}$Pt$_{0.5}$ alloy
compositional order indeed has a profound influence on the
magnetocrystalline anisotropy, especially in the way it breaks
cubic symmetry. In this paper, we describe our scheme in some 
detail and provide an in-depth study of this effect in another 
CoPt alloy and demonstrate an electronic origin of the 
enhancement of the magnetocrystalline anisotropy energy (MAE). 
Also, it is known experimentally that, upon ordering, the 
equiatomic CoPt alloy undergoes a modest tetragonal lattice 
distortion ($c/a$ = 0.98) which can also enhance the MAE. We 
have calculated the MAE of disordered fcc-Co$_{0.5}$Pt$_{0.5}$ 
alloy for different $c/a$ ratios and find that a 2\% 
tetragonalization would contribute only about 20\% of the 
observed MAE. This is a clear indication that in CoPt system, 
the enhancement of MAE is primarily due to compositional order.
This inference gets further credence from the fact that 
fcc-Co$_{0.25}$Pt$_{0.75}$ alloy, which retains its cubic 
lattice structure upon ordering, also shows an enhancement of 
MAE when compositional order is introduced.
 
We have studied the effects of compositional modulation on
the MAE of fcc-Co$_c$Pt$_{1-c}$ for $c$=0.25 and 0.5 alloys.
Thick films of these alloys are potential magneto-optical 
recording materials because they exhibit large perpendicular 
anisotropy,\cite{dw1,dw2,mm1,mm2,wg1,cjl} large magneto-optic 
Kerr effect signals compared to the Co/Pt multilayers and the 
currently used TbFeCo films for a large range of 
wavelengths \cite{dw1,dw2} (400-820 nm), have suitable Curie 
temperatures \cite{dt} ($\sim$700 K), high oxidation and 
corrosion resistance as well as being chemically stable and 
easy to manufacture. We investigate the observed ordered 
structures as well as some hitherto unfabricated structures of 
these alloys with same stoichiometry. This is pertinent now 
that it is possible to {\it tailor} compositionally modulated 
films to obtain better magneto-optic recording 
characteristics.\cite{rc,pp} We find that, compositional 
ordering enhances the size of MAE by some two orders of 
magnitude for all the compositions in addition to altering the 
magnetic easy axis in some cases. The easy axis in all the 
layered-ordered structures lies perpendicular to the 
layer-stacking. Of particular interest is the case of 
Co$_{0.25}$Pt$_{0.75}$. MAE for the perfect $ L1_2 $ ordered 
alloy is very small as in the homogeneously disordered alloy 
with the easy axis along the [111] direction. However, breaking
this symmetry by some kind of a directional chemical order and 
creating {\it internal interfaces} enhances the MAE by two 
orders of magnitude as well as making the easy axis 
perpendicular to the interfaces, in excellent agreement with 
the experimental observations.\cite{tat} By analyzing the
electronic structure of these alloys we find that the 
electrons near the Fermi surface, in particular, those
electrons around the X-points in the Brillouin zone, are 
largely responsible for the enhancement of MAE by
compositional ordering.

Using the same theoretical framework, we have also studied the
phenomenon of magnetic annealing in Ni$_{0.75}$Fe$_{0.25}$
permalloy, which develops uniaxial magnetic anisotropy when
annealed in a magnetic field.\cite{sc,etf} Because of its high 
permeability and low coercivity, the permalloy is a good soft 
magnet and can be used for low switching fields in sensors. 
Also, a large difference in the conductivity of majority and 
minority spins \cite{dmcn} makes it a good candidate for 
spin-valves, spin transistors and magnetic tunnel junctions. On
magnetic annealing, the permalloy develops directional chemical
order \cite{ln,st} which is responsible for the uniaxial
anisotropy. In the previous work \cite{our3} we showed for the
first time from first-principles theoretical calculations that 
magnetic annealing can indeed produce directional chemical 
order. Here we isolate the electronic origin of the effect.

The outline of the paper is as follows. In Sec. II, we present
the formulation. In Sec. III we provide some numerical and 
technical details. Afterwards, we present our results in Sec. 
IV and in Sec. V we investigate the electronic origin of the 
MAE enhancement as well as of the magnetic annealing. Finally, 
in Sec. VI we draw some conclusions. 

\section{FORMULATION}
Our theory is based on the relativistic spin-polarized local 
density functional theory \cite{ahm,akr,mvr} and its solution 
by the SPR-KKR-CPA method.\cite{he,ps} A detailed description
of the method for the homogeneously disordered ferromagnetic 
alloys is provided in Ref. \onlinecite{our1}, and will 
not be repeated. Here, we describe in detail our new framework 
to investigate the effects of compositional order, both short- 
and long-ranged, on the magnetocrystalline anisotropy (A summary
of the approach has already been described in our recent 
letter Ref. \onlinecite{our3}). In Sec. II A we give a brief 
outline of the theory of compositional order within the 
SPR-KKR-CPA formalism and in Sec. II B we describe the scheme 
for studying the effects of compositional order on the MAE 
together with the
theory of magnetic annealing. Also, in this section we describe
a scheme to investigate the electronic origin of the enhancement
of MAE upon ordering.

\subsection{Compositional order}
We consider a binary alloy $A_c B_{1-c}$ where the atoms 
are arranged on a fairly regular array of lattice sites. At
high temperatures the alloy is homogeneously disordered and
each site is occupied by an $A$- or $B$-type atom with 
probabilities $c$ and $(1-c)$ respectively. Below some
transition temperature, $T_c$, the system will either order or
phase separate. A compositionally modulated alloy can be 
described by a set of site-occupation variables $ \{ \xi_i \} $,
with $ \xi_i = 1(0) $ when the $i$-th site in the lattice is 
occupied by an $A(B)$-type atom. The thermodynamic average,
$ \langle \xi_i \rangle $, of the site-occupation variable
is the concentration $ c_i $ of an $A$-type atom at that site.
At high temperatures where the alloy is homogeneously 
disordered, $ c_i = c $ for all sites. When inhomogeneity
sets in below $ T_c $, the temperature-dependent inhomogeneous 
concentration fluctuations 
$ \{ \delta c_i \} = \{ c_i - c \} $ can be written as a 
superposition of static concentration waves,\cite{agk} i.e.,
\[
c_i = c + \frac12 \sum_{\bf q} \left[
           c_{\bf q} e^{i {\bf q} \cdot {\bf R}_i } +
           c_{\bf q}^\ast e^{-i {\bf q}
           \cdot {\bf R}_i } \right],
\]
where, $ c_{\bf q} $ are the amplitudes of the concentration
waves with wave-vectors {\bf q}, and $ {\bf R}_i $ are the
lattice positions. Usually only a few concentration waves are 
needed to describe a particular ordered structure. For example,
the CuAu-like $ L1_0 $ layered-ordered structure (Fig. 
\ref{figure1}) is set up by a
single concentration wave with $ c_{\bf q}=\frac12 $ and 
$ {\bf q}=(001) $, the [111]-layered CuPt-like $ L1_1 $
ordered structure is set up by a concentration wave with
$ c_{\bf q}=\frac12 $ and
$ {\bf q}=( \frac12 \frac12 \frac12 ) $, and
the Cu$_3$Au-like $ L1_2 $ ordered structure is set up by
three concentration waves of identical amplitude
$ c_{\bf q}=\frac14 $ and wave-vectors $ {\bf q}_1=(100) $,
$ {\bf q}_2=(010) $, and $ {\bf q}_3=(001) $ ({\bf q} is in 
units of $ \frac{2 \pi}{a}$, $ a $ being the lattice parameter).

The grand-potential for the interacting electrons in
an inhomogeneous alloy with composition $ \{ c_i \} $ and
magnetized along the direction {\bf e} at a finite
temperature $ T$ is given by,\cite{blg1,blg2,jbs}
\begin{eqnarray}
\Omega (\{ c_i \}; {\bf e} ) = \nu Z & - &
       \int \limits_{ - \infty}^\infty
           d \varepsilon \; f(\varepsilon , \nu )\;  
           N (\{ c_i \}, \varepsilon ; {\bf e} ) \nonumber \\
& + & \Omega_{DC} (\{ c_i \}; {\bf e} ), \label{eq:gp}
\end{eqnarray}
where, $ \nu $ is the chemical potential,
$ Z $ is the total valence
charge, $ f(\varepsilon , \nu ) $ is the Fermi factor,
$ N (\{ c_i \}, \varepsilon ; {\bf e} ) $ is the integrated
electronic density of states, and
$ \Omega_{DC} (\{ c_i \}; {\bf e} ) $ is the 
`double-counting' correction to the grand-potential.\cite{blg2}
The derivatives of the grand potential with respect to the
concentration variables give rise to a hierarchy of direct
correlation functions. In particular, the second derivative
evaluated at the equilibrium concentrations,
\[
S^{(2)}_{jk} ({\bf e}) = \left.
\frac{ \partial^2 \Omega (\{ c_i \}; {\bf e} ) }
     { \partial c_j \partial c_k}
 \right|_{ \{ c_i = c \} },
\]
is the Ornstein-Zernike direct correlation function for our
lattice model \cite{blg1,blg2,jbs,jbs3,ddj} (so called by 
the way of 
the close analogy with similar quantities defined for classical 
fluids \cite{re,jph}). These are related to the linear response
functions, $ \alpha_{ij} ({\bf e}) $, through \cite{jbs}
\begin{eqnarray}
\lefteqn{
\left[ 1 + \sum_k S^{(2)}_{ik} ({\bf e}) 
           \alpha_{ki} ({\bf e}) \right] \alpha_{ij} ({\bf e}) 
} \nonumber \\
& &  \phantom{mmmm} = \beta c ( 1 - c ) \left[ \delta_{ij} + 
      \sum_k S^{(2)}_{ik} ({\bf e}) 
      \alpha_{kj} ({\bf e}) \right],
\label{eq:lrf}
\end{eqnarray}
where $ \beta = (k_B T)^{-1} $, $ k_B $ being the Boltzmann
constant. The linear response functions, 
$ \alpha_{ij} ({\bf e}) $, describe the resulting 
concentration fluctuations, $ \{ \delta c_i \} $, which are
produced when a small inhomogeneous set of external chemical 
potentials, $ \{ \delta \nu_i \} $, is applied at all sites.
Via the fluctuation dissipation theorem these are proportional
to atomic pair-correlation functions, i.e. 
$ \alpha_{ij} = \beta [ \langle \xi_i \xi_j \rangle -
\langle \xi_i \rangle \langle \xi_j \rangle$]).
Upon taking the lattice Fourier transform of Eq. (\ref{eq:lrf})
we obtain a closed form of 
equations, \cite{blg1,blg2,jbs,jbs3,ddj}
\begin{equation}
\alpha ({\bf q},T; {\bf e}) = \frac{\beta c (1-c)}
   { 1 - \beta c (1-c) [ S^{(2)} ({\bf q}; {\bf e}) - 
  \lambda_c ]},
\label{eq:asro}
\end{equation}
where, the Onsager cavity correction $ \lambda_c $ is given 
by,\cite{jbs}
\[
\lambda_c = \frac{1}{\beta c (1-c)}
  \frac{1}{V_{BZ}} \int d {\bf q}' \;
  S^{(2)} ( {\bf q}'; {\bf e}) \: \alpha ( {\bf q}',T; {\bf e}).
\]
Here, $ \alpha ({\bf q},T) $, the lattice Fourier transform
of $ \alpha_{ij} $, are the Warren-Cowley atomic
short-range order (ASRO) parameters in the disordered phase.
The Onsager cavity correction in Eq. (\ref{eq:asro}) 
ensures that the spectral weight over the Brillouin zone is 
conserved,\cite{jbs,ddj,lo} so that, in other words, the 
diagonal part of the fluctuation dissipation theorem is 
honored, i.e. $ \alpha_{ii} = \beta c (1-c) $.

The spinodal transition 
temperature $ T_c $ below which the alloy orders
into a structure characterized by the concentration wave-vector
${\bf q}_{max}$ is determined by 
$ S^{(2)} ({\bf q}_{max}; {\bf e}) $, where $ {\bf q}_{max}$ is 
the value at which $S^{(2)} ({\bf q}; {\bf e})$ is maximal. 
Neglecting the Onsager cavity correction, 
we can write,\cite{blg1,blg2}
\[
T_c = \frac{c(1-c) S^{(2)} ({\bf q}_{max} ; {\bf e})}{k_B}.
\]

For the purpose of this study, we need only to consider the 
effects of compositional fluctuations in which charge 
rearrangement effects are neglected. 
Within the SPR-KKR-CPA scheme, a formula for
$ S^{(2)}_{jk} ({\bf e}) $ is obtained by using Lloyd
formula \cite{pl} for the integrated density of states in Eq.
(\ref{eq:gp}), 
\begin{eqnarray}
S^{(2)}_{jk} ({\bf e}) = & - & \frac{Im}{\pi} 
       \int \limits_{ - \infty}^\infty d \varepsilon \:
    f(\varepsilon , \nu )
 Tr \left[ \left\{ X^A ({\bf e}) - X^B ({\bf e}) 
        \right\} \frac{}{} \right. \nonumber \\
& & \phantom{mm}
 \times \sum_{m \ne j} \left. \frac{}{} \tau^{jm} ({\bf e}) \:
\Lambda^{mk} ({\bf e}) \: \tau^{mj} ({\bf e}) \right],
\label{eq:s2jk}
\end{eqnarray}
where,
\[
 X^{A(B)} ({\bf e}) = \left[ \left\{ t_{A(B)}^{-1} ({\bf e})
  - t_c^{-1} ({\bf e}) \right\}^{-1}+
\tau^{00} ({\bf e}) \right] ^{-1},
\]
and
\begin{eqnarray*}
\Lambda^{jk} ({\bf e}) & = & \delta_{jk} 
\{ X^A ({\bf e}) - X^B ({\bf e}) \} \nonumber \\
& - & X^A ({\bf e}) \sum_{m \ne j} \tau^{jm} ({\bf e}) \:
\Lambda^{mk} ({\bf e}) \: \tau^{mj} ({\bf e}) \:
 X^B ({\bf e}).
\end{eqnarray*}
Here  $ t_{A(B)} ({\bf e})$ and $ t_c ({\bf e}) $ are the 
t-matrices for electronic scattering from sites occupied by
$ A (B)$ atoms and CPA effective potentials respectively, and
$ \tau^{mj} ({\bf e}) $ are the path operator matrices
for the CPA effective medium in real space obtained by a lattice
Fourier transform of $ \tau ( {\bf k}; {\bf e}) $, where
\[
\tau ( {\bf k}; {\bf e}) = \left[ t_c^{-1} ({\bf e}) - 
 g ({\bf k}) \right]^{-1},
\]
$ g ({\bf k}) $ being the KKR structure constants 
matrix.\cite{jsf} Now taking the 
lattice Fourier transform of Eq. (\ref{eq:s2jk}), we get
\begin{eqnarray}
S^{(2)} ({\bf q}; {\bf e}) & = & - \frac{Im}{\pi}
  \int \limits_{ - \infty}^\infty d \varepsilon \;
   f(\varepsilon , \nu ) \nonumber \\
& \times & \sum_{L_1 L_2 L_3 L_4} \left[ \left\{
 X^A ({\bf e}) - X^B ( {\bf e}) \right\}_{ L_1 L_2 } 
\right. \phantom{mmmmm} \nonumber \\
 & & \left. \phantom{mi} \frac{}{} \times 
  I_{ L_2 L_3 ; L_4 L_1 } ( {\bf q} ; {\bf e}) \:
  \Lambda_{ L_3 L_4 } ({\bf q}; {\bf e} ) \right],
  \label{eq:s2q}
\end{eqnarray}
where,
\begin{eqnarray*}
\Lambda_{ L_1 L_2 } ( {\bf q}; {\bf e}) & = &
\left\{ X^A ({\bf e}) - X^B ({\bf e}) \right\}_{ L_1 L_2} \\
 & & - \sum_{L_3 L_4 L_5 L_6} \left[ \frac{}{}
  X^A_{ L_1 L_5} ({\bf e}) \:
 I_{ L_5 L_3 ; L_4 L_6 } ({\bf q}; {\bf e}) \right. \nonumber \\
& & \left. \phantom{mmmm} \frac{}{} \times 
X^B_{ L_6 L_2} ({\bf e}) \:
\Lambda_{ L_3 L_4 } ({\bf q} ; {\bf e}) \right],
\end{eqnarray*}
and,
\begin{eqnarray}
I_{ L_5 L_3 ; L_4 L_6 } ({\bf q}; {\bf e}) & = &
  \frac{1}{V_{BZ}} \int d {\bf k} \;
  \tau_{ L_5 L_3 } ( {\bf k} + {\bf q} ; {\bf e}) \:
  \tau_{ L_4 L_6 } ( {\bf k}; {\bf e}) \nonumber \\
& & - \tau_{ L_5 L_3 }^{00} ({\bf e}) \: 
\tau_{ L_4 L_6 }^{00} ({\bf e}). \label{eq:iq}
\end{eqnarray}
Experimentally, the instability of the disordered phase to
ordering can be observed in diffuse electron, x-ray, or
neutron scattering experiments. The experimentally measured
intensities are proportional to the ASRO parameter.
In previous works, the ASRO parameter $ \alpha ({\bf q},T)$ and
ordering temperature $ T_c$ have been calculated for many
alloys, both non-magnetic and ferromagnetic, and were compared
with diffuse x-ray and neutron scattering data.\cite{blg2,jbs}
However, in those studies, relativistic effects were largely
ignored. These studies have revealed that the electronic 
structure around the Fermi level is sometimes the driving force 
behind unusual compositional ordering in some 
alloys.\cite{blg1,jbs}
Schematically, $ S^{(2)} ({\bf q}; {\bf e})$ can be written
in terms of the convolution of Bloch spectral density
functions \cite{jsf} $ A_B ({\bf k}, \varepsilon  ; {\bf e}) $
as,\cite{blg1}
\begin{eqnarray}
S^{(2)} ({\bf q}; {\bf e}) & \sim & 
 \int \limits_{ - \infty}^\infty d \varepsilon
\! \int \limits_{ - \infty}^\infty d \varepsilon^\prime \:
  \frac{ f(\varepsilon , \nu ) - f(\varepsilon^\prime , \nu )}
       { \varepsilon - \varepsilon^\prime } \nonumber \\
& \times & \int d {\bf k} \;
  A_B ( {\bf k} + {\bf q} , \varepsilon ; {\bf e}) \:
  A_B ( {\bf k}, \varepsilon^\prime ; {\bf e}).
\label{eq:s2q3}
\end{eqnarray}
For perfectly ordered alloys, this reduces to a standard
susceptibility form.\cite{blg2} In some cases the principal 
contribution to
$ S^{(2)} ({\bf q}; {\bf e}) $ comes from energies close
to the Fermi level $ \nu $. Thus $ S^{(2)} ({\bf q}; {\bf e}) $
can be large around the Fermi energy in two ways. In the
conventional Fermi surface `nesting' mechanism \cite{blg1,blg2}
overlap takes place over extended regions of the reciprocal
space between almost parallel flat sheets of Fermi 
surface \cite{blg1} with spanning vector {\bf Q} as in 
Cu$_{0.75}$Pd$_{0.25}$. The structures based on this mechanism 
will tend to be long period or incommensurate structures. 
Alternatively, a large $ S^{(2)} ({\bf q}; {\bf e}) $ can 
result from a spanning vector connecting the van Hove 
singularities around high-symmetry points,\cite{jfc} as in 
Cu$_{0.5}$Pt$_{0.5}$. The structures based on this mechanism 
will tend to produce high symmetry structures with short 
periodicities since the spanning vector will be that which 
connects the high-symmetry points in the Brillouin zone.
We find similar Fermi surface effects contribute to the
enhancement of MAE by compositional order found in 
Co$_c$Pt$_{1-c}$ alloys, and propose that the mechanism is
widespread in other magnetic transition-metal alloys.

\subsection{Magnetocrystalline Anisotropy}
The MAE of the inhomogeneous alloy can be characterized by
the change in the electronic grand-potential arising from the
change in the magnetization direction. Thus, 
\[
K(\{ c_i \}) = \Omega (\{ c_i \}; {\bf e}_1 ) -
               \Omega (\{ c_i \}; {\bf e}_2 ),
\]
where, $ {\bf e}_1 $ and $ {\bf e}_2 $ are two magnetization
directions. We assume that the double-counting correction
$ \Omega_{DC} (\{ c_i \}; {\bf e}) $ is generally unaffected
by the change in the magnetization direction, and therefore,
only the first two terms of Eq. (\ref{eq:gp}) contribute to
the MAE,
\begin{eqnarray*}
K(\{ c_i \}) & = & ( \nu_1 - \nu_2 ) Z -
       \int \limits_{ - \infty}^\infty
           d \varepsilon \; f(\varepsilon , \nu_1 ) \;
           N (\{ c_i \}, \varepsilon ; {\bf e}_1 ) \\
     & + & \int \limits_{ - \infty}^\infty
           d \varepsilon \; f(\varepsilon , \nu_2 ) \;
           N (\{ c_i \}, \varepsilon ; {\bf e}_2 ),
\end{eqnarray*}
where $ \nu_1 $ and $ \nu_2 $ are the chemical potentials of the
system when the magnetization is along $ {\bf e}_1 $ and 
$ {\bf e}_2 $ directions respectively. The change in the 
chemical potential originates from a redistribution of the 
occupied energy bands in the Brillouin zone in the event of a 
change of magnetization direction. A Taylor expansion of
$ f(\varepsilon , \nu_2 ) $ about $ \nu_1 $ and some
algebra leads to, 
\begin{eqnarray*}
K(\{ c_i \}) & = & \int \limits_{ - \infty}^\infty d \varepsilon
   \; f(\varepsilon , \nu_1 ) \left[
   N (\{ c_i \}, \varepsilon ; {\bf e}_1 ) -
   N (\{ c_i \}, \varepsilon ; {\bf e}_2 ) \right] \\
   & + & O ( \nu_1 - \nu_2 )^2,
\end{eqnarray*}
Note that the effect of the small change in the chemical
potential on $ K(\{ c_i \}) $ is of second order in
$ ( \nu_1 - \nu_2 ) $, and can be
shown to be very small compared to the first term.\cite{our1}
We now expand $ K(\{ c_i \}) $ around $ K_{CPA} (c) $, the MAE
of the homogeneously disordered alloy $A_c B_{1-c}$,
\begin{eqnarray}
\lefteqn{K(\{ c_i \}) = K_{CPA} (c) + \sum_j \left.
\frac{ \partial K(\{ c_i \}) }{ \partial c_j }
\right|_{ \{ c_i = c \} } \delta c_j } \nonumber \\
 & & \phantom{mmm} + \frac12 \sum_{j,k} \left.
\frac{ \partial^2 K(\{ c_i \}) }{ \partial c_j \partial c_k }
\right|_{ \{ c_i = c \} } \delta c_j
\delta c_k + O ( \delta c )^3. \label{eq:k1}
\end{eqnarray}
It is clear from Eq. (\ref{eq:k1}) that we are considering 
only the effects of two-site correlations on the
electronic grand-potential and MAE. However, in principle it
is possible to include higher order correlations, but
computationally it will be prohibitive.

Within the SPR-KKR-CPA scheme, a formula for $ K_{CPA} (c) $ 
is obtained by using the Lloyd formula \cite{pl} for the 
integrated density of states,
\begin{eqnarray}
K_{CPA} (c) = & - & \frac{Im}{\pi} 
       \int \limits_{ - \infty}^\infty
    d \varepsilon \; f(\varepsilon , \nu_1 ) \left[
\frac{1}{V_{BZ}} \int d {\bf k} \right. \nonumber \\
& \times & \ln \| I + \left\{ t_c^{-1} ({\bf e}_2) -
 t_c^{-1} ({\bf e}_1) \right\}
\tau ({\bf k}; {\bf e}_1 ) \| \nonumber \\
& + &
c \; \left( \frac{}{} \ln \| D^A ({\bf e}_1) \| - 
   \ln \| D^A ({\bf e}_2) \| \right) \nonumber \\
& + & \left. 
(1-c) \; \left( \frac{}{} \ln \| D^B ({\bf e}_1) \| -
 \ln \| D^B ({\bf e}_2) \| \right) \right] \nonumber \\
& + & {\mathcal O} ( \nu_1 - \nu_2 )^2,
\label{eq:dealloyt}
\end{eqnarray}
where
\[
 D^{A(B)} ({\bf e}) = \left[I + \tau^{00} ({\bf e})
     \{ t_{A(B)}^{-1} ({\bf e}) - 
t_c^{-1} ({\bf e}) \} \right]^{-1}.
\]
Note that Eq. (\ref{eq:dealloyt}) 
is the finite temperature version of the expression of MAE of
disordered alloys given in Ref. \onlinecite{our1}. Now, the
variation of the MAE with respect to the change in the
concentration variables is,
\begin{eqnarray*}
\left. \frac{ \partial K(\{ c_i \}) }{ \partial c_j }
\right|_{ \{ c_i = c \} } = & - & \frac{Im}{\pi} 
       \int \limits_{ - \infty}^\infty d \varepsilon \;
   f(\varepsilon , \nu_1 ) \\
& \times & \left[ \frac{}{}
          \ln \| D^A ({\bf e}_1) \| -
          \ln \| D^B ({\bf e}_1) \| \right. \\
& & \left. -
          \ln \| D^A ({\bf e}_2) \| +
          \ln \| D^B ({\bf e}_2) \| \frac{}{} \right],
\end{eqnarray*}
which is independent of the site-index and so the second term in
Eq. (\ref{eq:k1}) vanishes if the number of $ A $ and $ B $
atoms in the alloy is to be conserved 
( $ \sum_j \delta c_j  = 0 $ ). Also, we have,
\[
\left.
\frac{ \partial^2 K(\{ c_i \}) }{ \partial c_j \partial c_k}
 \right|_{ \{ c_i = c \} } =
S^{(2)}_{jk} ({\bf e}_1) - S^{(2)}_{jk} ({\bf e}_2),
\]
where, $ S^{(2)}_{jk} ({\bf e}_{1(2)} )$ are the 
Ornstein-Zernike direct correlation functions \cite{blg1} when 
the magnetization is along $ {\bf e}_{1(2)} $.
Now taking the Fourier transform of Eq. (\ref{eq:k1}),
we get the MAE of the compositionally
modulated alloy with wave-vector {\bf q},
\begin{equation}
K({\bf q}) = K_{CPA} (c) + \frac12 \:
\vert c_{\bf q} \vert ^2 \: K^{(2)} ({\bf q}), \label{eq:k2}
\end{equation}
where
\begin{equation}
K^{(2)} ({\bf q}) = S^{(2)} ({\bf q}; {\bf e}_1) -
S^{(2)} ({\bf q}; {\bf e}_2) . \label{eq:k3}
\end{equation}

Eq. (\ref{eq:k2}) thus shows a direct relationship between the
type of compositional modulation and the MAE. Also, by
calculating $ S^{(2)} ({\bf q}; {\bf e} ) $ for different
{\bf q}-vectors, while keeping the magnetic field and
magnetization direction fixed, one can study the effect of an 
applied magnetic field on the compositional modulation of a
solid solution, and thus can describe the phenomenon of
magnetic annealing. In this case, the {\bf q}-vector for which
$ S^{(2)} ({\bf q}; {\bf e} ) $ is maximal will represent the
compositional modulation induced in the alloy when it is
annealed in the magnetic field.

In our previous studies on disordered
alloys \cite{our1,our2} we have found that the MAE originates
from the change in the electronic structure around the
Fermi energy caused by altering the magnetization direction.
The changed electronic structure around the Fermi surface caused
by the compositional order is thus the reason for the
enhancement of MAE in the ordered phase. To understand this
effect we undertake the following analysis. Using the same 
argument leading to Eq. (\ref{eq:s2q3}) we arrive at,
\begin{eqnarray*}
K^{(2)} ({\bf q}) & \sim &
  \int \limits_{ - \infty}^\infty d \varepsilon \!
  \int \limits_{ - \infty}^\infty d \varepsilon^\prime \:
  \frac{ f(\varepsilon , \nu ) - f(\varepsilon^\prime , \nu )}
       { \varepsilon - \varepsilon^\prime } \\
& \times & \int d {\bf k} \;
 \left[ A_B ( {\bf k} + {\bf q} , \varepsilon ; {\bf e}_1 ) \:
  A_B ( {\bf k}, \varepsilon^\prime ; {\bf e}_1 ) \right. \\
& & \phantom{mm} \left. -
A_B ( {\bf k} + {\bf q} , \varepsilon ; {\bf e}_2 ) \:
  A_B ( {\bf k}, \varepsilon^\prime ; {\bf e}_2 ) \right]
\end{eqnarray*}
which can be rewritten as,
\begin{eqnarray}
K^{(2)} ({\bf q}) & \sim & \frac12
  \int \limits_{ - \infty}^\infty d \varepsilon 
  \int \limits_{ - \infty}^\infty d \varepsilon^\prime \:
  \frac{ f(\varepsilon , \nu ) - f(\varepsilon^\prime , \nu )}
       { \varepsilon - \varepsilon^\prime } \nonumber \\
& \times & \int d {\bf k} \; \left[
  \Sigma ( {\bf k} + {\bf q} , \varepsilon ) \:
  \Delta ( {\bf k}, \varepsilon^\prime ) \right. \nonumber \\
& & \left. \phantom{mm} +
  \Sigma ( {\bf k} , \varepsilon^\prime ) \:
  \Delta ( {\bf k} + {\bf q} , \varepsilon ) \right]
  \label{eq:k4}
\end{eqnarray}
where,
\begin{equation}
\Sigma ( {\bf k} , \varepsilon ) = 
      A_B ( {\bf k} , \varepsilon ; {\bf e}_1) +
      A_B ( {\bf k} , \varepsilon ; {\bf e}_2)
      \label{eq:sigma}
\end{equation}
and
\begin{equation}
\Delta ( {\bf k} , \varepsilon) = 
      A_B ( {\bf k} , \varepsilon ; {\bf e}_1) -
      A_B ( {\bf k} , \varepsilon ; {\bf e}_2).
      \label{eq:delta}
\end{equation}
The principal 
contributions to $ K^{(2)} ({\bf q}) $ will come
from energies close to the Fermi energy $ \nu $. Thus, 
$ K^{(2)} ({\bf q}) $ will be large for {\bf q}-vectors
corresponding to the spanning vectors $ {\bf q} = {\bf Q} $
connecting the peaks in
$ \Sigma ( {\bf k} , \varepsilon ) $ and
$ \Delta ( {\bf k} , \varepsilon ) $ for $ \varepsilon $
close to the Fermi energy. Therefore, a close examination
of $ \Sigma ( {\bf k}, \nu ) $ and $ \Delta ( {\bf k} , \nu ) $
will give an idea as to which composition modulations will
produce a large MAE.

\section{COMPUTATIONAL DETAILS}
In this section we discuss the technical details in the 
calculation of MAE of compositionally modulated alloys. In a 
homogeneously disordered alloy, the MAE is of the order of 
$\mu$eV and the MAE of the ordered systems can be as high as 
0.5meV. But the MAE is still several orders of magnitude smaller
than the band energies as well as the total energy. For this
reason the calculation of MAE needs great care. In an earlier
publication \cite{our1} we presented a robust scheme to 
calculate the MAE, where we proposed that one should calculate
the difference in the total energies for the two magnetization
directions directly rather than subtracting the two total
energies calculated separately. We included a detailed 
discussion on the computational and technical aspects of 
solving the SPR-KKR-CPA equations as well as the calculation of
MAE of the disordered alloys and do not repeat that 
discussion here. Instead, we will focus only on the calculation
of $ S^{(2)} ( {\bf q}; {\bf e}) $ and $ K^{(2)} ( {\bf q}) $. 

The integration over the energies involved in Eqs.
(\ref{eq:s2q}) and (\ref{eq:k3}) is carried out using a complex 
contour because in the complex plane, the integrand becomes 
a smooth function of both energy as well as {\bf k} and 
consequently we need fewer energy points as well as fewer 
{\bf k}-points to obtain an accurate integral.\cite{rz}
In our calculation, we have used a rectangular box contour as 
described in detail in Ref. \onlinecite{our1}.

The most demanding part of the whole calculation is the
evaluation of the convolution $ I ({\bf q}; {\bf e}) $ given by
Eq. (\ref{eq:iq}). Owing to the form of the integrand 
the integration has to be done using the full Brillouin zone.
We performed the Brillouin zone integration by using the
adaptive grid method \cite{eb} which has been found to be very
efficient and accurate. In this method one can preset the
level of accuracy of the integration by supplying a tolerance
parameter $ \epsilon $. The number of {\bf k}-points used
in the integration thus depends directly on  $ \epsilon $.
In the present work, the integration in Eq. (\ref{eq:iq}) is
carried out with $ \epsilon = 10^{-6} $ which means that
values of $ S^{(2)} ({\bf q}; {\bf e}) $ 
which are of the order of
0.1 eV are accurate up to 0.1 $ \mu $eV. To achieve such level
of accuracy, we had to sample a large number of {\bf k}-points
in the Brillouin zone. Typically, in our calculations, we needed
around 35,000 {\bf k}-points in the full Brillouin zone
for energies with an imaginary part of 0.5 Ry. When the energy
was 0.0001 Ry above the real axis (and that is the closest point
to the real axis, we need) we required as many as 3 million
{\bf k}-points in the full Brillouin zone for the same level
of accuracy. Furthermore, we have calculated
$ S^{(2)} ({\bf q}; {\bf e}_1) $ and
$ S^{(2)} ({\bf q}; {\bf e}_2) $
simultaneously ensuring that they are calculated on the same 
grid of {\bf k}-points, and thus cancelling out the
systematic errors if there are any. Therefore, we claim that 
the values of $ K ({\bf q}) $ are accurate to within 
0.1 $ \mu $eV.

\section{RESULTS AND DISCUSSION}

\subsection{Co$_c$Pt$_{1-c}$ alloy}
We have studied Co$_c$Pt$_{1-c}$ alloys for $c$=0.5 and 0.25. 
It is well known that disordered fcc-Co$_{0.5}$Pt$_{0.5}$ 
undergoes a phase transformation into a CuAu-type $ L1_0 $ 
ordered tetragonal structure \cite{gh,mh} with a $ c/a $ ratio 
of 0.98 below 1100 K, and Co$_{0.25}$Pt$_{0.75}$ orders into a 
Cu$_3$Au-type $ L1_2 $ cubic structure \cite{cl} below 960 K.
Thus, in case of the equiatomic composition, the 
tetragonalization of the lattice which also lowers the symmetry
could also be responsible for MAE enhancement, whereas in
Co$_{0.25}$Pt$_{0.75}$ there is no such additional effect,
because the lattice remains cubic even in the ordered phase. 
With this in mind we calculated the MAE of disordered
volume-conserving face-centered-tetragonal Co$_{0.5}$Pt$_{0.5}$ 
alloy as a function of the $c/a$ ratio. The results are 
presented in Fig. \ref{figure2}. We point out that in these
calculations we have used atomic-sphere approximation for
the single-site potentials and that the potentials and the
Fermi energy are those of the disordered 
fcc-Co$_{0.5}$Pt$_{0.5}$ alloy. To estimate the magnitude of 
the effect of tetragonalization we have `frozen' these 
potentials as the $c/a$ ratio has been altered. We observe
that the MAE is a monotonically decreasing function of the
$c/a$ ratio, and that it is positive for $c/a < 1.0 $ and 
negative for $c/a > 1.0$. This is consistent with the
experimental observations that the magnetostriction 
constant, $ \lambda_{001} $, is positive \cite{sh,jaa} for 
the disordered Co$_{0.5}$Pt$_{0.5}$ alloy. Because, Freeman 
{\it et al} \cite{ajf} have shown that $ \lambda_{001} $ is
proportional to the rate of change of MAE with respect to the
$ c/a $ ratio, and that $ \lambda_{001} $ has the opposite sign
to that of the later (Note that, there is a sign difference
in our definition of MAE and that of Freeman {\it et al}).
Therefore, our results are in qualitative agreement with the
experimental observations. The MAE at the experimental value
of $ c/a $ (0.98) is about 25 $\mu eV$ which is less than
20\% of the experimentally observed MAE. Most importantly,
the sign of MAE for this value of $c/a$ is positive which
means that the magnetic easy axis is not along the [001]
direction ($c$-axis) in direct contradiction to the experimental
observations. Consequently, we conclude that in the ordered
alloys of Co and Pt lattice distortion is not the major factor
in determining the MAE rather it is the compositional order
which is primarily responsible for the large MAE.

\begin{figure}
\center{\psfig{figure=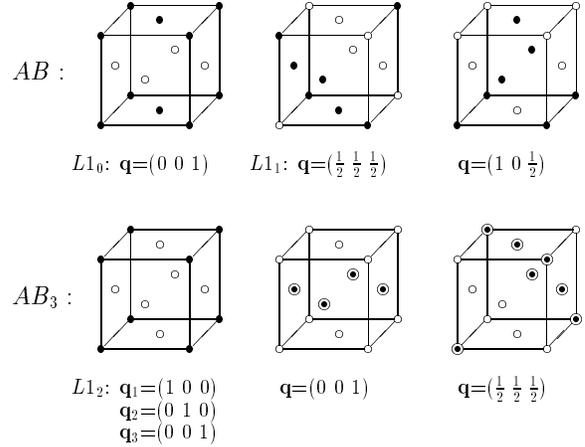,height=2.75in}}
\caption{Some ordered structures and their representative
concentration wave-vectors. For the $AB$-type stoichiometry
{\bf q}=(001) and {\bf q}=($\frac12 \frac12 \frac12 $) generate
respectively the CuAu-type $ L1_0 $ layered ordered structure
with layers perpendicular to the [001] direction and
the CuPt-type $ L1_1 $ layered-ordered structure with layers
perpendicular to the [111] direction and 
{\bf q}=($ 1 0 \frac12$) generates a layered structure with 
planes of an ordered structure of $ A $ and $ B $ atoms stacked 
along the [001] direction. For the $ AB_3$-type composition, 
a combination of {\bf q}$_1$=(100), {\bf q}$_2$=(010), and
{\bf q}$_3$=(001) generates the Cu$_3$Au-type $ L1_2 $
ordered structure. For this composition, a single wave-vector 
{\bf q}=(001) generates a superstructure of alternating 
monolayers of pure $B$ atoms and disordered $A_{0.5}B_{0.5}$
perpendicular to the [001] direction. Similarly, 
{\bf q}=($\frac12 \frac12 \frac12 $) generates a superstructure
of monolayers of pure $B$ atoms and disordered $A_{0.5}B_{0.5}$
perpendicular to the [111] direction. The full circles denote 
$A$ atoms, open circles denote $B$ atoms, and a full circle 
inscribed by an open circle denotes a CPA effective atom 
$A_{0.5}B_{0.5}$.}
\label{figure1}
\end{figure}

\begin{figure}
\center{\psfig{figure=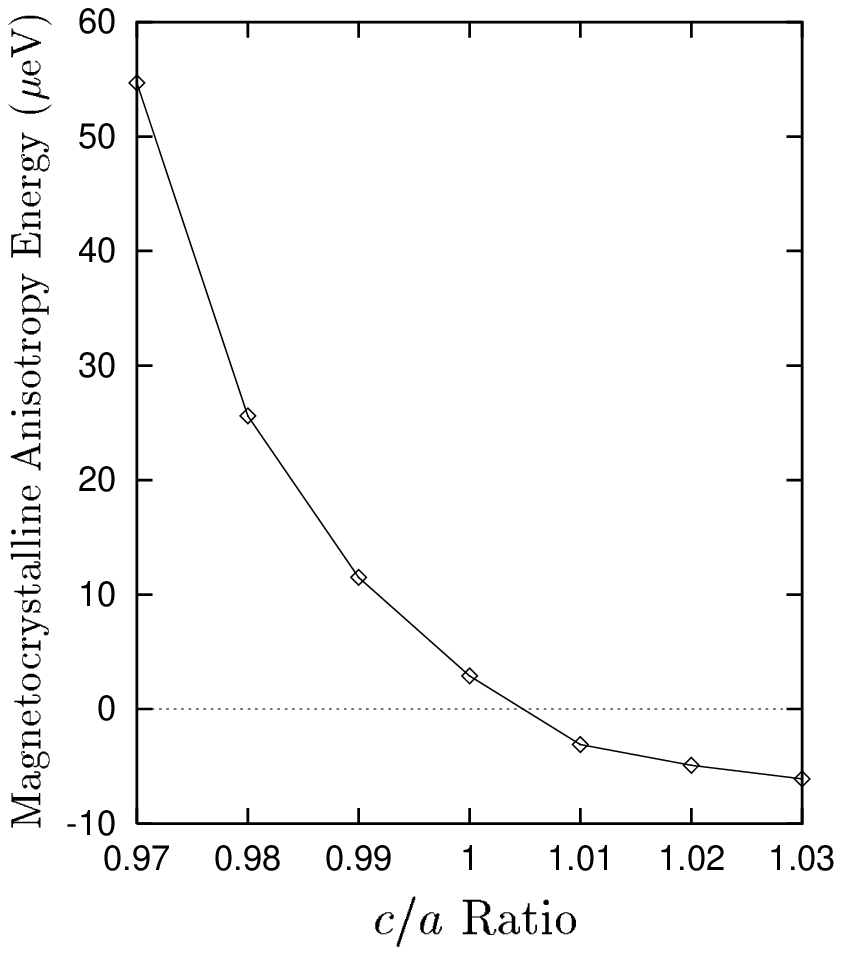,height=3.20in}}
\caption{Magnetocrystalline anisotropy energy (MAE) of 
disordered volume-conserving face-centered tetragonal 
({\it fct}) Co$_{0.5}$Pt$_{0.5}$ alloy as a function of the
$ c/a $ ratio.}
\label{figure2}
\end{figure}

Now we present the effect of compositional order on the 
magnetocrystalline anisotropy. In our studies, we explore the
observed equilibrium ordered structures as well as some 
hypothetical ordered structures of these alloys keeping the
stoichiometry constant.
The results are summarized in Tables \ref{table1} and
\ref{table2}. First we discuss Co$_{0.5}$Pt$_{0.5}$. We note 
that, $ S^{(2)} ({\bf q}) $ is maximum for the $ L1_0 $ 
structure, implying that the alloy would order into this 
structure at 1360 K as it is cooled from high temperature in 
good agreement with experiment (experimental value of the 
ordering temperature is 1100 K). From Table \ref{table2} we 
note that for $ {\bf q}=(001) $ and $ {\bf q}=(100) $ which 
correspond to CuAu-like $ L1_0 $ ordered structure, with Co and 
Pt layers stacked along the [001] and [100] directions 
respectively, the direction of spontaneous magnetization is 
along the [001] and [100] directions respectively in excellent 
agreement with experiment. Also, the MAE (58.6 $\mu $eV) is 
comparable to the experimental value \cite{gh,bz,ram} 
($\sim 130 \mu $eV) as well as to the calculated value of the
$ L1_0 $-ordered tetragonal CoPt alloy.\cite{ivs} In the 
CuPt-type $ L1_1 $ layered structure set up by 
$ {\bf q}=( \frac12 \frac12 \frac12 ) $, the equilibrium 
direction of magnetization is along the [111] direction of the 
crystal, i.e. again perpendicular to the layer-stacking. 
The structure
set up by concentration waves with $ {\bf q}=(1 0 \frac12) $ and
$ (0 1 \frac12) $ is also a [001]-oriented layered structure, 
but the layers are not alternately pure Co and Pt planes, rather
they are layers of CoPt in a particular order 
(Fig. \ref{figure1}). Even in this case, we note that the 
magnetization is perpendicular to the layered structure. 
Similarly, for $ {\bf q}=(\frac12 0 1) $ and
$ (\frac12 1 0) $ where the planes are stacked along the 
[100] direction the magnetization is also along the [100] 
direction.

In case of Co$_{0.25}$Pt$_{0.75}$, $ S^{(2)} ({\bf q}) $ has
a maxima for {\bf q}=(100), (010), and (001) implying that
a $ L1_2 $ ordered structure is favored with a transition
temperature of 935 K in excellent agreement with the
experimental value of 960 K (Ref. \onlinecite{cl}). Note that, 
a combination of three wave-vectors $ {\bf q}_1 = (100)$, 
$ {\bf q}_2 = (010)$, and $ {\bf q}_3 = (001)$ generates the 
isotropic $ L1_2 $ ordering, while a single wave-vector, for 
example, $ {\bf q}_1 = (001)$ generates a layered structure 
with directional compositional 
ordering along the [001] direction. This is a superstructure
consisting of alternating monolayers of pure Pt and
Co$_{0.5}$Pt$_{0.5}$, as depicted in Fig. \ref{figure1}. 
In this structure, therefore, there are no out-of-plane
Co-Co bonds, only in-plane Co-Co bonds which can produce only 
in-plane Co-Co nearest neighbor pairs. We find that for this 
structure, the easy axis is along the [001] direction, 
{\it i.e.} perpendicular to the layers, and the MAE is also 
quite large ($\sim 84 \mu$eV). In a recent 
experiment,\cite{si1} it was found that, the [001]-textured 
thick films of Co$_{0.25}$Pt$_{0.75}$ alloy deposited at 670 K 
do have this type of structure, {\it i.e.} there are stacks of 
Pt and Co$_{0.5}$Pt$_{0.5}$ monolayers perpendicular to the 
[001] direction, and these films exhibit PMA. It should be
emphasized, however, for the 
perfect $ L1_2 $ structure, where all the three wave-vectors, 
namely, $ {\bf q}_1 = (100)$, $ {\bf q}_2 = (010)$, and 
$ {\bf q}_3 = (001)$ contribute, the easy axis is along the 
[111] direction and the MAE is very small, comparable to that
of the disordered alloy.
Another [111]-stacked layered structure is
generated by $ {\bf q} = ( \frac12 \frac12 \frac12 ) $ in which
there are alternating planes of pure Pt and 
Co$_{0.5}$Pt$_{0.5}$. In this case also the easy axis is 
perpendicular to the layer stacking (i.e. along [111]) and the
MAE is 23.4 $\mu$eV. We are 
not aware of any experimental results on the bulk ordered 
Co$_{0.25}$Pt$_{0.75}$ system. However, it is reported that
[111]-textured thick films grown around 690 K having 
anisotropic compositional order exhibit large uniaxial 
anisotropy \cite{mm1,mm2,pwr} whereas films deposited around 
800 K with a $ L1_2 $-type isotropic chemical order exhibit no 
anisotropy. These observations are clearly in good agreement 
with our results.

The above observations suggest that a large MAE and an easy 
axis perpendicular to the layer-stacking is a result of 
existence of in-plane Co-Co bonds and out-of-plane Co-Pt bonds.
It has been observed experimentally that in these films there 
are indeed more Co-Co bonds in-plane and Co-Pt bonds 
out-of-plane which produce {\it internal interfaces} analogous 
to Co/Pt multilayers.\cite{tat} Because of the out-of-plane 
orientation of Co-Pt pairs a hybridization between the Co atoms
with large magnetic moment and Pt atoms with strong spin-orbit 
coupling induces anisotropies in the 3d and 5d orbital moments,
the out-of-plane components of the orbital moments become 
larger than the in-plane components, as has been observed in 
the x-ray magnetic circular dichroism measurements in the 
CoPt$_3$ films.\cite{wg1,wg2} The induced anisotropy in the 
orbital moments directs the spin moment into a perpendicular
alignment through spin-orbit coupling, thus overcoming the
in-plane shape anisotropy due to the spin-spin dipole
interaction.\cite{dw3}

\subsection{Magnetic annealing of Ni$_{0.75}$Fe$_{0.25}$ alloy}
When Ni$_{0.75}$Fe$_{0.25}$ permalloy is annealed in a magnetic
field, an uniaxial magnetic anisotropy is induced depending on
the direction of the applied field with respect to the 
crystallographic axes.\cite{sc,etf} This phenomenon, known as 
magnetic annealing, is interpreted as the creation of directional
chemical order in the material.\cite{ln,st} A recent study
based on magneto-optic Kerr effect measurements \cite{fs} also 
reveals that magnetic annealing can induce uniaxial anisotropy. 
Previous studies based on non-relativistic electronic structure 
calculations \cite{jbs2} and Monte Carlo simulations \cite{mzd}
have shown that in this system compositional and magnetic 
ordering have a large influence on each other. However, to 
describe magnetic annealing a fully relativistic treatment is
needed. We produce the first quantitative description of 
magnetic annealing from {\it ab initio} electronic structure 
calculations in Ni$_{0.75}$Fe$_{0.25}$. Our results are 
summarized in Table \ref{table3}. We calculate 
$S^{(2)} ({\bf q},{\bf e})$ for the permalloy in an applied 
magnetic field of strength 600 Oe (same as used in the 
experiment \cite{sc}) oriented along ${\bf e}=$ [001], [111],
and [100] directions of the crystal. We find that when the 
magnetic field is along [001] (column 2) $ S^{(2)} ({\bf q}) $
is maximum for {\bf q}=(001) confirming that ordering is 
favored along the direction of applied field. In this case, the 
compositional structure is a layered structure along the [001] 
direction comprising of alternate pure Ni layers and disordered 
Ni$_{0.5}$Fe$_{0.5}$ layers. Noting that the measured intensity 
in a scattering experiment is proportional to the ASRO 
parameter $ \alpha ( {\bf q} ) $, we present the ASRO parameters
calculated at 722 K (1 K above the ordering temperature) for 
different {\bf q}-vectors in Table \ref{table4}. We note that 
when the applied magnetic field is along the [001] direction, 
$ \alpha ( {\bf q} ) $ for {\bf q}=(001) is 40-50\% larger
than that at other {\bf q}-vectors. Therefore, when the alloy 
is annealed in the magnetic field, the superlattice spot in the 
measured intensity at {\bf q}=(001) will be 40-50\% more 
intense than that at {\bf q}=(100) at a temperature 1 K above 
the transition temperature. When the magnetic field is applied 
along the [100] direction, a similar structure perpendicular to 
[100] direction is favored. However, when the applied field is 
along [111] direction (column 6) all the three orderings, 
namely, (100), (010), and (001) are favored equally. Thus, in 
this case, we will get a Cu$_3$Au-type $ L1_2 $ ordering. The 
calculated transition temperature 721 K is in good agreement 
with the experimental value \cite{jo} of 820 K.

We have also calculated the MAE of Ni$_{0.75}$Fe$_{0.25}$
permalloy for different ordered structures. We found that
the MAE of the disordered phase is very small (less than
$ 0.1 \mu$eV) and in case of directional chemical ordering
it is of the order of $ \mu$eV, an increase by an order of 
magnitude. However, in case of perfect ordering, which is 
$ L1_2 $, the MAE becomes very small as in the disordered 
phase. This is consistent with experimental observations
of Chikazumi \cite{sc} and Ferguson.\cite{etf}

\section{ELECTRONIC ORIGIN OF MAGNETIC ANISOTROPY}
In elemental solids and the disordered alloys the MAE originates
from a redistribution of electronic states around the Fermi 
level caused by the change in the magnetization 
direction.\cite{our1} The electronic structure of the 
disordered phase around the Fermi level is also partly
responsible for the tendency to compositional order in some 
alloys.\cite{blg1,jbs} The enhancement of MAE in the 
compositionally modulated alloys is also related to the 
electronic structure around the Fermi level in the disordered 
phase. Consequently, 
to understand the electronic origin of large MAE in the 
ordered phase, in the following we analyze the Bloch spectral 
function of the disordered Co$_{0.5}$Pt$_{0.5}$ alloy along
the lines as depicted in Eq. (\ref{eq:k4}).

\begin{figure}
\center{\psfig{figure=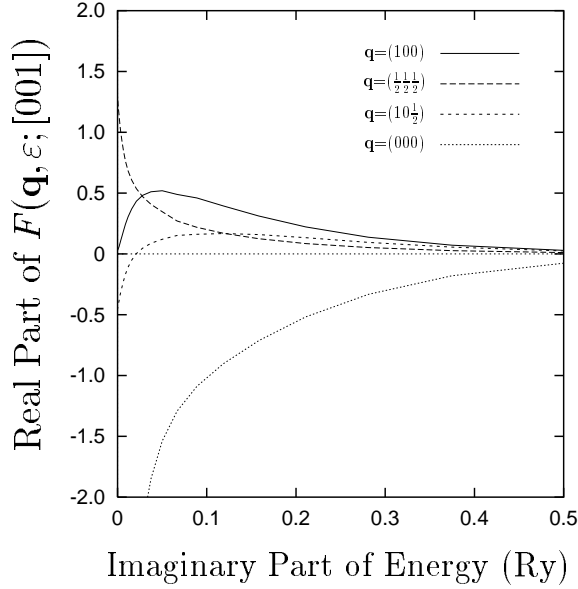,height=3.60in}}
\caption{Real part of $ F ( {\bf q}, \varepsilon, [001])$,
as given by Eq. (\ref{eq:fqe}), for
{\bf q}=(100) (full line), ($\frac12 \frac12 \frac12$) 
(long-dashed line), (10$\frac12$) (short-dashed line), and
(000) (dotted line) at complex energies along the imaginary 
axis perpendicular to the Fermi level for disordered
Co$_{0.5}$Pt$_{0.5}$ alloy.}
\label{figure3}
\end{figure}

First we consider the ordering tendency. As presented in Table
\ref{table1}, our calculation predicts $ L1_0 $-type order
in Co$_{0.5}$Pt$_{0.5}$ which is in excellent agreement with
experiment. The quantity $ S^{(2)} ({\bf q} ; {\bf e}) $ 
given by Eq. (\ref{eq:s2q}) can be rewritten as,
\[
S^{(2)} ({\bf q}; {\bf e}) = - \frac{Im}{\pi}
  \int \limits_{ - \infty}^\infty d \varepsilon \;
   f(\varepsilon , \nu ) 
   F ({\bf q}, \varepsilon ; {\bf e})
\]
where,
\begin{eqnarray}
F ({\bf q}, \varepsilon ; {\bf e}) & = &
\sum_{L_1 L_2 L_3 L_4} \left[ \left\{
 X^A ({\bf e}) - X^B ( {\bf e}) \right\}_{ L_1 L_2 } 
\right. \nonumber \\
 & & \phantom{m} \left. \frac{}{} \times 
  I_{ L_2 L_3 ; L_4 L_1 } ( {\bf q} ; {\bf e}) \:
  \Lambda_{ L_3 L_4 } ({\bf q}; {\bf e} ) \right].
  \label{eq:fqe}
\end{eqnarray}
$ S^{(2)} ({\bf q}) $ is an integrated quantity and can also be
written in terms of a sum \cite{kw} over Matsubara 
frequencies,\cite{gdm} $ \omega_n = ( 2n+1) \pi k_B T $,
\[
S^{(2)} ({\bf q}; {\bf e}) = 2 k_B T \sum_n 
   Re \left[ F ({\bf q}, \nu + i \omega_n ; {\bf e})
    \right].
\]
In Fig. \ref{figure3} 
we show a plot of $ F ({\bf q}, \varepsilon ; [001]) $ 
calculated for energies along the imaginary axis perpendicular 
to the Fermi level for {\bf q}-vectors (000), (001), 
($ \frac12 \frac12 \frac12$) and ($ 1 0 \frac12 $). The area 
under these curves is indicative of the strength of 
$ S^{(2)} ({\bf q}; [001]) $ for these {\bf q}-vectors and is 
obviously greatest for {\bf q}=(100).

\begin{figure}
\center{\psfig{figure=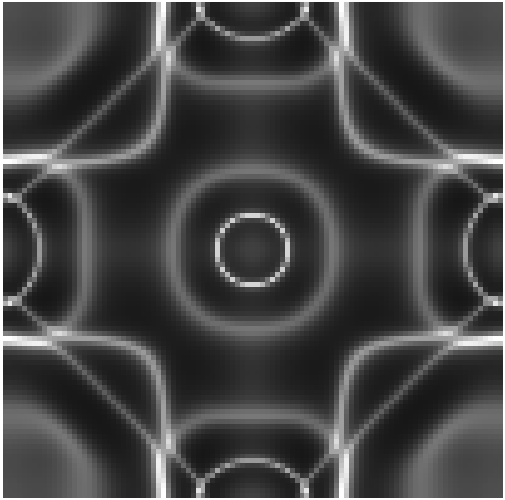,width=3.0in}}
\caption{Density-plot of the Bloch spectral density function of 
disordered Co$_{0.5}$Pt$_{0.5}$ alloy at the Fermi energy in the
(001) plane ($ k_z = 0$). The $ \Gamma$-points are at the 
corners and the X-points are at the center of the edges. White 
areas indicate relatively larger density of states.}
\label{figure4}
\end{figure}

In special cases the electronic structure near the Fermi 
surface can explain some unusual ordering 
tendencies. One famous example is a nesting of the Fermi
surface \cite{blg1,blg2} which takes place over 
extended regions of the reciprocal space between almost 
parallel sheets of Fermi surfaces as in Cu$_{0.75}$Pd$_{0.25}$
and gives rise to its incommensurate ordering tendency.
However, presence of van Hove singularities can give rise to a 
somewhat different kind of mechanism of ordering in which the 
spanning vector couples only the regions around two 
high-symmetry points \cite{jfc} as in the case of CuPt. In 
Fig. \ref{figure4} we show a density-plot of the Bloch spectral 
density function of the disordered Co$_{0.5}$Pt$_{0.5}$ alloy 
in the (001) plane ($ k_z = 0$). In this figure the 
$ \Gamma$-points are at the corners and the X-points are at the 
center of the edges. White areas indicate relatively larger 
density of states. The cross-section of the Fermi surface sheets 
are seen quite clearly. In this figure we do not observe any 
van Hove singularities at the symmetry points, the ordering
tendency coming instead from states over a wide energy range.
However, we do 
notice that, there are large densities of states around the 
X-points as well as around the (110) points.

\begin{figure}
\center{\psfig{figure=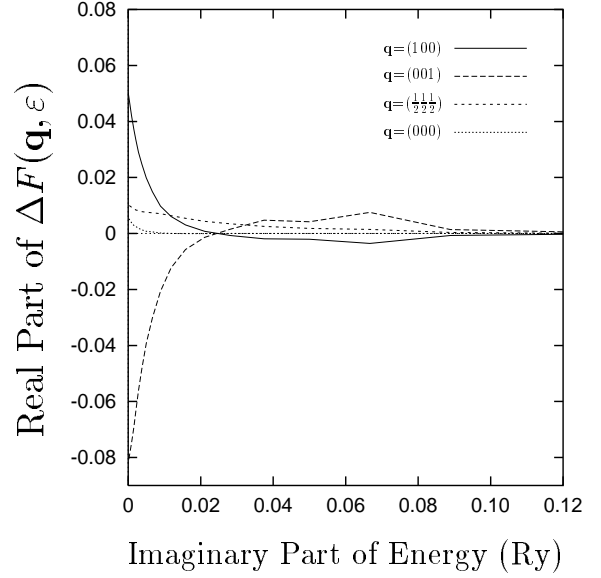,height=3.50in}}
\caption{Real part of $ \Delta F ( {\bf q}, \varepsilon)$,
as defined in Eq. (\ref{eq:delf}), with 
$ {\bf e}_1 = [001] $ and $ {\bf e}_2 = [111] $ for 
{\bf q}=(100) (full line), (001) (long-dashed line),
($\frac12 \frac12 \frac12$) (short-dashed line), and
(000) (dotted line) at complex energies along the imaginary 
axis perpendicular to the Fermi level for Co$_{0.5}$Pt$_{0.5}$
alloy. Note the different scales on the $y$-axes of this 
figure and Fig. \ref{figure3}.}
\label{figure5}
\end{figure}

Now we discuss the enhancement of MAE. Again, the quantity
$ K^{(2)} ({\bf q}) $ given by Eq. (\ref{eq:k3}) is an 
integrated quantity which can be written as a sum of 
contributions evaluated at the Matsubara frequencies,\cite{kw}
\begin{equation}
K^{(2)} ({\bf q}) = 2 k_B T \sum_n 
   Re \left[ \Delta F ({\bf q}, \nu + i \omega_n )
    \right] \label{eq:k2q2}
\end{equation}
where
\begin{equation}
\Delta F ({\bf q}, \varepsilon) = 
    F ({\bf q}, \varepsilon; {\bf e}_1 ) - 
    F ({\bf q}, \varepsilon ; {\bf e}_2 ).
\label{eq:delf}
\end{equation}
In Fig. \ref{figure5} we show a plot of 
$ \Delta F ({\bf q}, \varepsilon) $ with $ {\bf e}_1 = [001] $
and $ {\bf e}_2 = [111] $ calculated for energies along the 
imaginary axis perpendicular to the Fermi level for 
{\bf q}-vectors (000), (001), ($ \frac12 \frac12 \frac12$) and 
($ 1 0 \frac12 $). The area under these curves is indicative of 
the magnitude of $ K^{(2)} ({\bf q}) $ for these {\bf q}-vectors.
The striking feature is that the principal contributions are 
near the real axis indicating that the Fermi surface plays 
the dominant role in the enhancement of the MAE as well as the 
direction of easy magnetization. {\it We propose that such Fermi
surface contributions are widespread factors in the 
enhancement of MAE by compositional order in many magnetic
alloys}.

\begin{figure}
\center{\psfig{figure=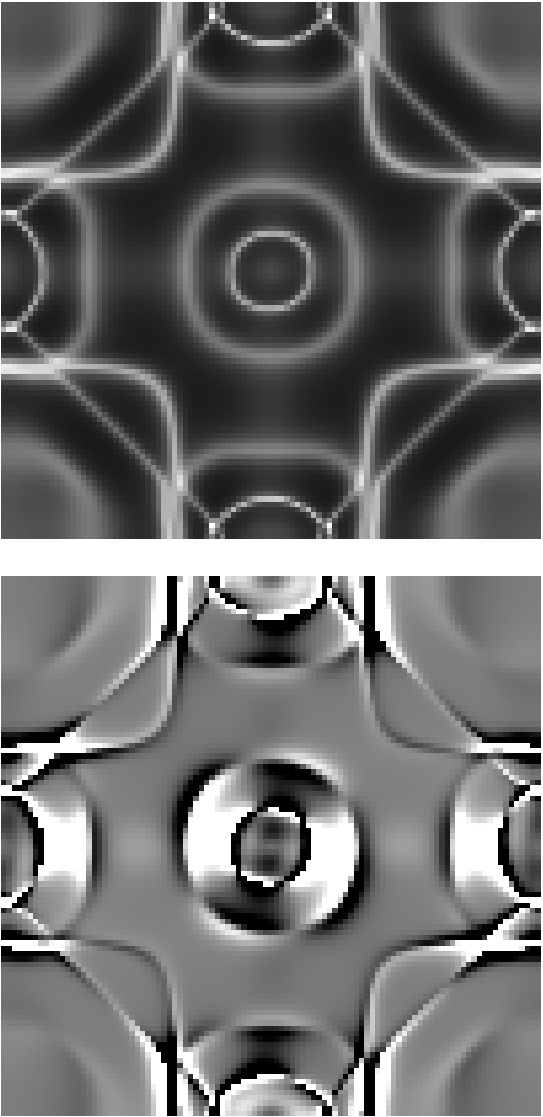,width=3.0in}}
\caption{Density-plot of the sum over two different
magnetization directions of the Bloch spectral
density functions $ \Sigma ( {\bf q} , \nu ) $ (top) and
the difference of the Bloch spectral density functions
$ \Delta ( {\bf q} , \nu ) $ (bottom) as defined in 
Eqs. (\ref{eq:sigma}) and (\ref{eq:delta}), for disordered 
Co$_{0.5}$Pt$_{0.5}$ alloy at the Fermi energy in the (010) 
plane ($ k_y = 0$). The $ \Gamma$-points are at the 
corners and the X-points are at the center of the edges. White 
areas indicate relatively larger density of states. In 
the bottom figure the dark areas indicate negative values
rather than zero.}
\label{figure6}
\end{figure}

The origin of the enhancement of MAE and the role of the Fermi
surface can be easily understood by analyzing the sum and
difference of Bloch spectral density functions for the two
magnetization directions as outlined in Eq. (\ref{eq:k4}).
Obviously, we are looking for the spanning vector {\bf Q} which 
connects the high density regions in 
$ \Sigma ( {\bf q}, \nu ) $, the sum of the Bloch spectral
density functions for two magnetisation directions,
and $ \Delta ( {\bf q} , \nu) $, the difference of the Bloch
spectral density functions for two magnetisation directions,
as defined in Eqs. (\ref{eq:sigma})
and (\ref{eq:delta}). In Fig. \ref{figure6} we show a
density-plot of $ \Sigma ( {\bf q}, \nu ) $ and 
$ \Delta ( {\bf q} , \nu) $ of the disordered 
Co$_{0.5}$Pt$_{0.5}$ alloy at the Fermi energy in the (010) 
plane ($ k_y = 0$). In this figure the $ \Gamma$-points are at 
the corners and the X-points are at the center of the edges. In 
the density-plot of $ \Sigma ( {\bf q}, \nu ) $ the white areas
indicate relatively larger values and the most dark areas 
indicate zero values, whereas in the density-plot of 
$ \Delta ( {\bf q} , \nu) $ the most white (most dark) areas 
indicate large positive (negative) values and the grey areas 
indicate intermediate values. As in case of the Bloch spectral 
density function, we do not observe large values at the symmetry
points. The analysis is more complicated for other
{\bf q}-vectors but we can conclude that the
enhancement of MAE for concentration wave-vectors near the
Brillouin zone boundary is indeed due to large values of
$ \Sigma ( {\bf q}, \nu ) $ and $ \Delta ( {\bf q} , \nu) $
at the Brillouin zone boundary.

\begin{figure}
\center{\psfig{figure=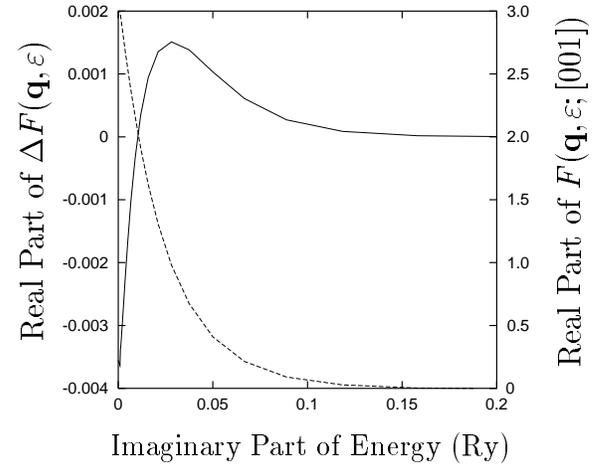,height=3.00in}}
\caption{Real part of $ \Delta F ( {\bf q}, \varepsilon)$ 
as defined in Eq. (\ref{eq:delf}) with $ {\bf e}_1 = [001] $ and 
$ {\bf e}_2 = [100] $ (full line, left scale) and real part of 
$ F ( {\bf q}, \varepsilon ; [001]) $, as defined in
Eq. (\ref{eq:fqe}) for {\bf q}=(001)
(dashed line, right scale) at complex energies along the 
imaginary axis perpendicular to the Fermi level for 
Ni$_{0.75}$Fe$_{0.25}$ alloy.}
\label{figure7}
\end{figure}

Now we discuss the electronic origin of magnetic annealing
effect in Ni$_{0.75}$Fe$_{0.25}$ alloy. We note from Table
\ref{table3} that the difference between 
$ S^{(2)} ( {\bf q}; [001] ) $ and $ S^{(2)} ( {\bf q}; [100])$
for different {\bf q}-vectors is quite small, but as depicted
in Table \ref{table4} it is large enough to be observed
by diffuse x-ray, electron and neutron 
scattering experiments. Because of the small
difference in these quantities, in Fig. \ref{figure7} we plot
the real part of $ \Delta F ({\bf q}, \varepsilon) $ which is
related to the difference between 
$ S^{(2)} ( {\bf q}; [001] ) $ and $ S^{(2)} ( {\bf q}; [100])$ 
for {\bf q}=(001) for energies along the imaginary axis 
perpendicular to the Fermi level (left-hand scale). We also plot 
$ F ({\bf q}, \varepsilon; [001] $ which is related to
$ S^{(2)} ( {\bf q}; [001] ) $ for {\bf q}=(001) for the same
energies (right-hand scale). We note that, these two quantities 
peak near the Fermi energy, i.e. when the imaginary part of the
energy is very small. This is an indication that this process
is also governed by the electrons near the Fermi surface. 

\section{CONCLUSIONS}
We have presented the details of our {\it ab initio} theory 
of the connection of
magnetocrystalline anisotropy of ferromagnetic alloys with the 
compositional order within the SPR-KKR-CPA scheme.
This theory has been applied
to fcc-Co$_c$Pt$_{1-c}$ alloys. We found that when 
cooled from a high temperature, fcc-Co$_{0.5}$Pt$_{0.5}$ tends to
order into $L1_0$ layered-ordered structure around 1360 K and 
fcc-Co$_{0.25}$Pt$_{0.75}$ tends to order into $ L1_2 $ 
structure around 935 K, in good agreement with experimental 
observations. Also, we found that in $L1_0$ ordered CoPt the 
spontaneous magnetization is along the [001] direction which 
assumes the $c$-axis of the tetragonal structure, in excellent 
agreement with experiment. The magnitude of the MAE is also 
close to the experimental value. In some of the other 
hypothetical layered-ordered structures with same stoichiometry
also the magnetic easy axis lies perpendicular to 
the layer stacking, with the [111]-stacked CuPt-like $ L1_1 $ 
structure having the largest MAE. In the $ L1_2 $ ordered 
CoPt$_3$ the magnetic easy axis is along the [111] direction 
which is the same as in its disordered counterpart. The MAE is 
also not very large. However, in case of directional ordering 
along either of the [100], [010] or [001] direction the
easy axis is along the ordering with greatly enhanced MAE. 
Finally, by analyzing the electronic structure of the 
disordered alloy near the Fermi energy we have found that the 
Fermi surface plays the dominant role in the enhancement of MAE.
Within the same theoretical framework we have also been able to 
explain the appearance of directional chemical order in 
Ni$_{0.75}$Fe$_{0.25}$ when it is annealed in an applied 
magnetic field and linked that also to the alloy's Fermi
surface. As a last remark, we look forward to future work in
which the effects of compositional structure are fully
incorporated into micromagnetic modeling of transition
metallic materials via {\it ab initio} electronic structure
calculations.

\section*{ACKNOWLEDGMENTS}
We thank B.L. Gyorffy for many helpful discussions and
encouragement. This research is supported by the Engineering 
and Physical Sciences Research Council (UK), National Science 
Foundation (USA), and the Training and Mobility of Researchers 
Network on ``Electronic structure calculation of materials
properties and processes for industry and basic sciences''.
We also thank the computing center CSAR at Manchester University
as part of the calculations were performed on their Cray T3E
machine.

\begin{table}
\caption{Direct correlation function
$ S^{(2)} ({\bf q}; [001])$ for different {\bf q}-vectors for 
Co$_c$Pt$_{1-c}$ alloys (the respective ordered structures are 
shown in Fig. \ref{figure1}).}
\begin{tabular}{ddddd}
   &     &            & $ S^{(2)} ({\bf q}; [001]) $ 
       & $ T_c $ \\
$ c $ & {\bf q} & Structure  & (eV)  &  (K) \\
\tableline
& (000)                         & Clustering & -1.51 &       \\
& (001)                         & $ L1_0 $   &  0.47 & 1360  \\
0.5 & (100)                     & $ L1_0 $   &  0.47 & 1360  \\
& ($ \frac12 \frac12 \frac12 $) & $L1_1  $   &  0.29 &       \\
& ($ 1 0 \frac12 $)             &            &  0.19 &       \\
\tableline
& (100)                         & $ L1_2 $   &  0.43 &  935  \\
& (010)                         & $ L1_2 $   &  0.43 &  935  \\
0.25 & (001)                    & $ L1_2 $   &  0.43 &  935  \\
& ($ \frac12 \frac12 \frac12 $) &            &  0.22 &       \\ 
& ($ 1 \frac12 0 $)             & $DO_{22}$  &  0.19 &       \\
& ($ \frac12 0 1 $)             & $DO_{22}$  &  0.19 &       \\
\end{tabular}
\label{table1}
\end{table}

\onecolumn

\begin{table}
\caption{Magnetocrystalline anisotropy energy
$ K ({\bf q}) $ for several compositionally modulated
 Co$_c$Pt$_{1-c}$
alloys characterized by different {\bf q}-vectors (the
respective ordered structures are shown in 
Fig. \ref{figure1}). Here
$ K ({\bf q}) $ are calculated with respect to the reference 
system which has the magnetization along the [001] direction 
(i.e. $ {\bf e_1} = [001] $) of the crystal. Thus, when
$ K ({\bf q}) < 0 $ the easy axis is along [001] and when
$ K ({\bf q}) > 0 $ the easy axis is along ${\bf e}_2$.}
\begin{tabular}{dddddd}
 &         &       & ${\bf e}_2=[111]$ & ${\bf e}_2=[100]$ &  \\
 & {\bf q} & Structure & $ K({\bf q}) $ & $ K({\bf q}) $ &
                     Easy Axis \\
 $ c $ & & & ($ \mu $eV) & ($ \mu $eV) & \\
\tableline
& (001)    &  $ L1_0 $            &  -58.6 & -105.6 & [001] \\
& (100)    &  $ L1_0 $            &   39.6 &  105.9 & [100] \\
0.5 & ($\frac12 \frac12 \frac12$) & $L1_1 $ &
                      152.0 & 0.0 & [111] \\
&($10\frac12$)& [001]-Layered & -158.7&-236.5& [001] \\
&($\frac12 01$)& [100]-Layered & 85.0 & 236.3 & [100] \\
\tableline
& (100)    & [100]-Layered &   30.5 &  83.6  & [100] \\
& (010)    & [010]-Layered &   30.5 &   0.0  &  -    \\
& (001)    & [001]-Layered &  -53.0 & -83.6  & [001] \\
0.25 & ($ \frac12 \frac12 \frac12 $) & 
     [111]-Layered &  23.4 &  0.0   & [111] \\
& (100),(010),(001) &  $ L1_2 $   &  8.0   &  0.0   & [111] \\
& ($ 1 \frac12 0 $) &            & -52.4 &  0.0   & [001] \\
& ($ 0 1 \frac12 $) &            & 100.2 & 152.5  & [100] \\
\end{tabular}
\label{table2}
\end{table}

\twocolumn

\begin{table}
\caption{Direct correlation function
$ S^{(2)} ({\bf q}; {\bf e})$ for different {\bf q}-vectors for 
Ni$_{0.75}$Fe$_{0.25}$ alloy (the respective ordered structures
are shown in Fig. \ref{figure1}).}
\begin{tabular}{ddddddd}
 & \multicolumn{2}{c}{{\bf e}=[001]}
 & \multicolumn{2}{c}{{\bf e}=[100]}
 & \multicolumn{2}{c}{{\bf e}=[111]} \\
 {\bf q} & $ S^{(2)} ({\bf q})$ & $ T_c $
         & $ S^{(2)} ({\bf q})$ & $ T_c $
         & $ S^{(2)} ({\bf q})$ & $ T_c $ \\
 & (meV) & (K) & (meV) & (K) & (meV) & (K) \\
\tableline
(100)   &  330.880 & & 331.158 & 720.9 & 330.973 & 720.5  \\
(010)   &  330.880 & & 330.880 & & 330.973 & 720.5  \\
(001)   &  331.158 & 720.9 & 330.880 & & 330.973 & 720.5  \\
($ \frac12 \frac12 \frac12 $) & -99.291 &
                                          & -99.291 &
                                          & -99.475 &  \\
($ 1 0 \frac12 $) & 99.284 & & 99.096 & & 99.222 & \\
($ \frac12 0 1 $) & 99.096 & & 99.285 & & 99.222 & \\
\end{tabular}
\label{table3}
\end{table}

\begin{table}
\caption{Atomic short-range order (ASRO) parameter
$ \alpha ({\bf q}; {\bf e})$ calculated at temperature 1 K above
the ordering temperature 721 K at few high-symmetry 
{\bf q}-vectors for Ni$_{0.75}$Fe$_{0.25}$ alloy (the 
respective ordered structures are shown in Fig. \ref{figure1}).}
\begin{tabular}{dddd}
 {\bf q} & {\bf e}=[001] & {\bf e}=[100] & {\bf e}=[111] \\
\tableline
(100)                         & 1509.7 & 2601.4 & 1756.2 \\
(010)                         & 1509.7 & 1509.7 & 1756.2 \\
(001)                         & 2601.4 & 1509.7 & 1756.2 \\
($ \frac12 \frac12 \frac12 $) &    2.3 &    2.3 &    2.3 \\
($ 1 0 \frac12 $)             &    4.3 &    4.3 &    4.3 \\
($ \frac12 0 1 $)             &    4.3 &    4.3 &    4.3 \\
\end{tabular}
\label{table4}
\end{table}

\end{document}